\documentclass{Interspeech}

\interspeechcameraready

\title{Recreating Neural Activity During Speech Production with Language and Speech Model Embeddings}

\author[affiliation={1}]{Owais Mujtaba}{Khanday}
\author[affiliation={2}]{Pablo Rodríguez San}{Esteban}
\author[affiliation={3}]{Zubair Ahmad}{Lone}
\author[affiliation={2}]{Marc }{Ouellet}
\author[affiliation={1}]{Jose A.}{Gonzalez-Lopez}

\affiliation{CITIC}{University of Granada, Granada, 18071}{Spain}
\affiliation{Dept. of Experimental Psychology}{University of Granada, Granada, 18071}{Spain}
\affiliation{GITAM, Vishakapatnam}{Andhra Pradesh, 530045}{India}
\email{\{owaismujtaba, prodriguez, moullet, joseangel\}@ugr.es, zlone@gitam.edu}

\usepackage{comment}

\begin{document}

\maketitle

\begin{abstract}

     Understanding how neural activity encodes speech and language production is a fundamental challenge in neuroscience and artificial intelligence. This study investigates whether embeddings from large-scale, self-supervised language and speech models can effectively reconstruct high-gamma neural activity characteristics, key indicators of cortical processing, recorded during speech production. We leverage pre-trained embeddings from deep learning models trained on linguistic and acoustic data to represent high-level speech features and map them onto these high-gamma signals. We analyze the extent to which these embeddings preserve the spatio-temporal dynamics of brain activity. Reconstructed neural signals are evaluated against high-gamma ground-truth activity using correlation metrics and signal reconstruction quality assessments. The results indicate that high-gamma activity can be effectively reconstructed using large language and speech model embeddings in all study participants, generating Pearson's correlation coefficients ranging from 0.79 to 0.99.
\end{abstract}

\section{Introduction} \label{sec:intro}
    Speech neuroprostheses, also known as speech brain-computer interfaces (BCI), are a class of assistive technology aimed at restoring or increasing communication abilities for individuals with severe speech and motor impairments. These devices decode neural activity associated with speech production and translate it into text \cite{Herff2015, Moses2021, Metzger2022, Willett2023, Silva2024bilingual, Card2024}, synthesized speech \cite{Anumanchipalli2019, Angrick2019, Herff2019, Angrick2021, Angrick2024, Khanday2025}, or other forms of communication, such as control parameters for 3D-model facial animation \cite{Metzger2023}. The potential applications of these devices are numerous \cite{Gonzalez2020}, including the promise to restore communication to individuals with conditions such as amyotrophic lateral sclerosis (ALS) \cite{Luo2023, Willett2023, Angrick2024, Card2024, Vansteensel2024} or other severe speech disorders caused by brain damage (e.g. brain-stem stroke) \cite{Moses2021, Metzger2022}. During the past decade, this area has experienced significant growth, driven by advances in neuroscience, deep learning, and neural interface technologies. Recent studies have demonstrated significant progress in decoding speech from neural signals, with applications ranging from text generation to real-time speech synthesis (see \cite{Puffay2023, Silva2024, Chang2024} for recent reviews on the topic). This interdisciplinary field, which bridges neuroscience and speech and language technology, has garnered increasing interest from both the scientific community and the public, as it holds the promise of restoring one of the most fundamental human abilities: our capacity to communicate.

    Most prior research on speech neuroprostheses has relied on invasive neural recording techniques, such as electrocorticography (ECoG) or microelectrode arrays, due to their superior signal quality compared to non-invasive methods like electroencephalography (EEG). Among these, ECoG, which involves placing electrodes directly on the brain's surface, has been particularly effective in capturing high-resolution neural activity related to speech production \cite{Herff2015, Anumanchipalli2019, Chen2024}. However, these methods are limited in their ability to access deeper brain structures, which may also play a critical role in speech and language processing. In this study, we investigate the potential of stereo EEG (sEEG), a less commonly used but highly promising sensing technology for speech neuroprostheses \cite{Herff2020}. Unlike ECoG, sEEG employs depth electrodes that can record neural activity from both cortical and subcortical regions, offering a more comprehensive view of the brain's speech network. Although sEEG has primarily been utilized in clinical settings for epilepsy monitoring, its potential for BCI applications, particularly in speech decoding, remains underexplored. This work aims to contribute to the limited but growing body of research on sEEG's applicability in speech neuroprostheses \cite{Angrick2021, Kohler2022, Angrick2022, Khanday2025}.
    
    In parallel to the above studies, recent neuroscience research has revealed a striking similarity between neural activity in the human brain during language-related tasks and the internal states of large-scale, self-supervised speech and language models. Models such as Wav2Vec 2.0 \cite{Baevski2020wav2vec} and GPT-2 \cite{Radford2019} have demonstrated a close alignment with brain activity during both speech perception and production, indicating that they encode linguistically relevant information in ways that parallel human neural processing \cite{Millet2022, Tuckute2023, Chen2024self}. These findings suggest that these models may serve as powerful tools for advancing neuroscience research, offering new insights into the neural basis of speech and language, as well as for developing more effective speech neuroprostheses.

    In this work, we aim to bridge these two lines of research by exploring whether embeddings derived from pre-trained language and speech models can effectively reconstruct neural activity captured during speech production using sEEG. To achieve this, we utilize embeddings from large deep-learning models trained on text and audio to represent high-level linguistic features, which are then mapped onto sEEG signals. We assess how well these embeddings preserve the spatio-temporal dynamics of brain activity by comparing the reconstructed neural signals to ground truth recordings, employing correlation metrics and signal reconstruction quality evaluations. Ultimately, this work aims to provide novel insights into the neural mechanisms underlying speech production and to evaluate the potential of large deep learning models to advance neuroscience research and the development of speech neuroprostheses.
    
\section{Method} \label{sec:method}
    \begin{figure}[t]
        \centering
        \includegraphics[width=0.95 \linewidth]{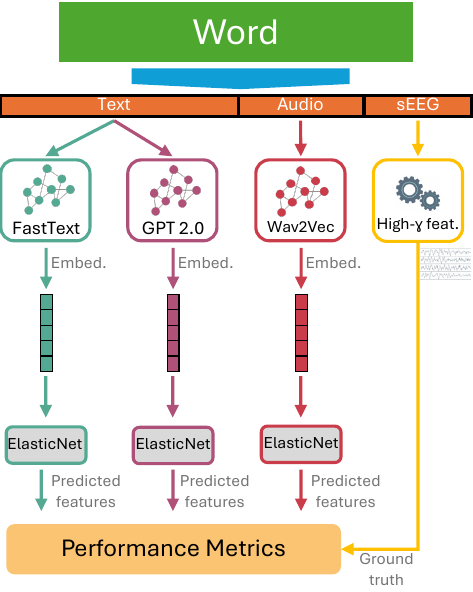}
        \caption{Methodology overview.}
        \label{fig:analysis_diagram}
    \end{figure}

    Fig. \ref{fig:analysis_diagram} presents a block diagram summarizing our methodology for this work. We used high-gamma features extracted from sEEG signals as the target neural features that were predicted from the embeddings computed by large, self-supervised audio and text models. These recordings were obtained from 10 participants implanted with depth electrodes while reading words aloud. The embeddings computed by the models were mapped onto the high-gamma features using an Elastic Net regression model \cite{tay2023elastic}, applied per subject. We evaluated the embeddings from three models: FastText \cite{fasttext} and GPT-2.0 \cite{Radford2019} for textual representations, and Wav2Vec 2.0 \cite{Baevski2020wav2vec} for audio representations. Finally, the reconstructed high-gamma features were compared to the ground-truth features extracted from the sEEG signals using objective metrics to assess reconstruction quality. The following sections describe the analysis performed in more detail.

    \subsection{Dataset} \label{ssec:dataset}
        We used a publicly available \cite{Verwoert2022} dataset comprising simultaneous recordings of audio and sEEG from 10 Dutch participants (5 male and 5 female; mean age: 32 years) diagnosed with pharmacoresistant epilepsy. As part of their clinical treatment, participants were implanted with platinum-iridium sEEG electrode shafts (Microdeep intracerebral electrodes; Dixi Medical, Besançon, France). Each electrode shaft had a diameter of 0.8 mm and contained between 5 and 18 electrode contacts, resulting in an average of 131.7 implanted electrodes per participant (SD = 49.92). The placement of the electrodes was determined solely based on clinical requirements, with a focus on regions such as the superior temporal sulcus, hippocampus, and inferior parietal gyrus. Due to individualized clinical needs, the number and location of electrodes varied between participants, providing sparse but widespread coverage of the cortical and subcortical regions. 
        
        During the experiment, participants read aloud a list of 100 words from the Dutch IFA corpus \cite{ifaCorpus}.The words were presented in a randomized order to avoid any content-based or alphabetical bias. The sEEG signals were recorded at sampling rates of either 2048 Hz or 1024 Hz, synchronized with the participants’ speech, which was captured at 48 kHz. For subsequent analysis, the sEEG recordings were downsampled to 1024 Hz, while the speech signals were downsampled to 16 kHz. To ensure participant anonymity, the audio recordings were pseudo-anonymized using a pitch-shifting technique.
    
    \subsection{Signal processing} \label{sec:preprocessing}
        sEEG signals were first linearly detrended to remove baseline drifts and linear trends, enhancing signal stability. To mitigate power-line interference, a notch filter was applied using two 4th-order IIR bandstop filters to attenuate the first two harmonics of the 50 Hz noise. The filters were applied bidirectionally to prevent phase distortion. From the cleaned signals, high-gamma power features (70–170 Hz) were extracted, as this frequency range has been shown to encode critical information about speech and language processing in the brain \cite{Angrick2024, Moses2021, Herff2015}. This was achieved by applying a 4th-order, non-causal Butterworth bandpass filter, followed by envelope extraction using the Hilbert transform. The resulting analytic signals were segmented into 50 ms windows with a 10 ms frameshift, and signal power was computed for each window to obtain the high-gamma features. 
        Implementation details can be found at \url{https://github.com/owaismujtaba/llm_brain_representations}

    \subsection{Word-based embeddings} \label{sec:wordembedding}
        We evaluated two text-based models, FastText and GPT-2.0, to extract linguistic embeddings from words spoken by the participants.
        
        FastText \cite{fasttext} is a classical word-embedding model that extends the bag-of-words approach by incorporating subword information via n-grams. It represents words as sums of vectorized character n-grams, enhancing its ability to handle morphological variations and out-of-vocabulary (OOV) words. Pre-trained on large-scale corpora like Common Crawl and Wikipedia, it provides static word-level embeddings with a fixed dimensionality of $d_{FastText}=300$. 
        
        GPT-2.0 \cite{Radford2019}, a transformer-based autoregressive model, generates contextualized word embeddings using self-attention mechanisms. Trained on 40GB of diverse internet text, it tokenizes input using Byte-Pair Encoding (BPE), improving subword processing and handling of OOV words. We extract embeddings from the penultimate transformer layer, yielding token-level representations with a dimensionality of $d_{GPT2}=(T, 768)$, where $T$ is the number of BPE tokens per word. To obtain a fixed-dimensional word representation, we compute the final embedding as the average of the embeddings of its $T$ tokens.

    \subsection{Audio-based embeddings}
        For computing audio-based embeddings, we employed the Wav2Vec 2.0 XLS-R model with 300 million parameters, initially described in \cite{babu2021xls}. This model is trained in a self-supervised manner on 436k hours of unlabeled speech from 128 languages, including datasets such as VoxPopuli, MLS, CommonVoice, BABEL, and VoxLingua107. The raw speech signal, sampled at 16 kHz, is processed through a convolutional feature encoder that extracts vector representations. These are subsequently passed through a 24-layer Transformer network, generating contextualized embeddings of the speech signal.
        
        For each 2-second audio segment corresponding to a spoken word, the model outputs an embedding tensor of shape $d_{Wav2Vec}=(24, 99, 1024)$, where 24 represents the number of Transformer layers, 99 corresponds to the number of computed feature timesteps, and 1024 is the feature dimension. In our setup, we specifically retain the embeddings from the last layer of the model, leveraging the most refined representations of the speech signal. Further, principal component analysis (PCA) was used to reduce the dimensionality of the embedding space, retaining 90\% of the variance. Given the multilingual training of XLS-R, this model is well-suited for computing embeddings of the Dutch words spoken by the participants.
    
    \subsection{EEG reconstruction forward model}
        To establish a relationship between the linguistc/audio embeddings and the high-gamma feaures computed from the sEEG signals, we trained an ElasticNet model \cite{tay2023elastic} for each type of embedding and participant (30 models in total). ElasticNet is a regularized, linear-regression technique that combines both L1 (Lasso) and L2 (Ridge) penalties, balancing effective feature selection and handling multicollinearity. The optimization objective is given by:  
    
        \begin{equation}
            \underset{\beta}{\min} \quad ||Y - X W||^2_2 + \lambda_1 ||W||_1 + \lambda_2 ||W||_2^2,
        \end{equation}
       
        \noindent where \( X \in \mathbb{R}^{n \times d} \) represent the input embeddings, \( n \) is the number of words in the training set, and \( d \) is the embedding dimension. The neural features (high-gamma) are denoted by \( Y \in \mathbb{R}^{n \times m} \), where \( m \) is the number of neural recording channels, and $W$ is the weight matrix of linear regression. To enforce sparsity in the learned mapping, L1 and L2 penalties are introduced and controlled by the parameters $\lambda_1$  and $\lambda_2$.
        

        \subsection{Evaluation and performance metrics}
        We used a leave-one-out (LOO) evaluation scheme to assess how well the high-gamma features can be predicted from text and audio embeddings. In each iteration, the ElasticNet model was trained on data from 99 words, with the remaining word serving as the test sample. This process was repeated 100 times to ensure that every word was used as a test case exactly once.
        
        To assess reconstruction accuracy, we employed two widely used objective metrics in statistics: the Pearson correlation coefficient (PCC) and the coefficient of determination ($R^2$). PCC measures how well the predicted signals ($\hat{y}$) preserve their temporal structure with respect to the ground-truth ones ($y$) :  
\begin{equation}
    \text{PCC} = \frac{\sum_{i=1}^{N} (y_i - \bar{y}) (\hat{y}_i - \bar{\hat{y}})}
    {\sqrt{\sum_{i=1}^{N} (y_i - \bar{y})^2} \sqrt{\sum_{i=1}^{N} (\hat{y}_i - \bar{\hat{y}})^2}}
\end{equation}    
    
    The $R^2$ coefficient, on the other hand, quantifies the proportion of variance in the neural data explained by the model predictions:  
\begin{equation}
    R^2 = 1 - \frac{\sum_{i=1}^{N} (y_i - \hat{y}_i)^2}{\sum_{i=1}^{N} (y_i - \bar{y})^2}
\end{equation}  
\noindent where a higher $R^2$ score indicates better reconstruction performance.  


\section{Results}
    \subsection{Performance of text-based embeddings}
            \begin{figure}[t]  
            \centering 
            \footnotesize
            \includegraphics[width=\linewidth]{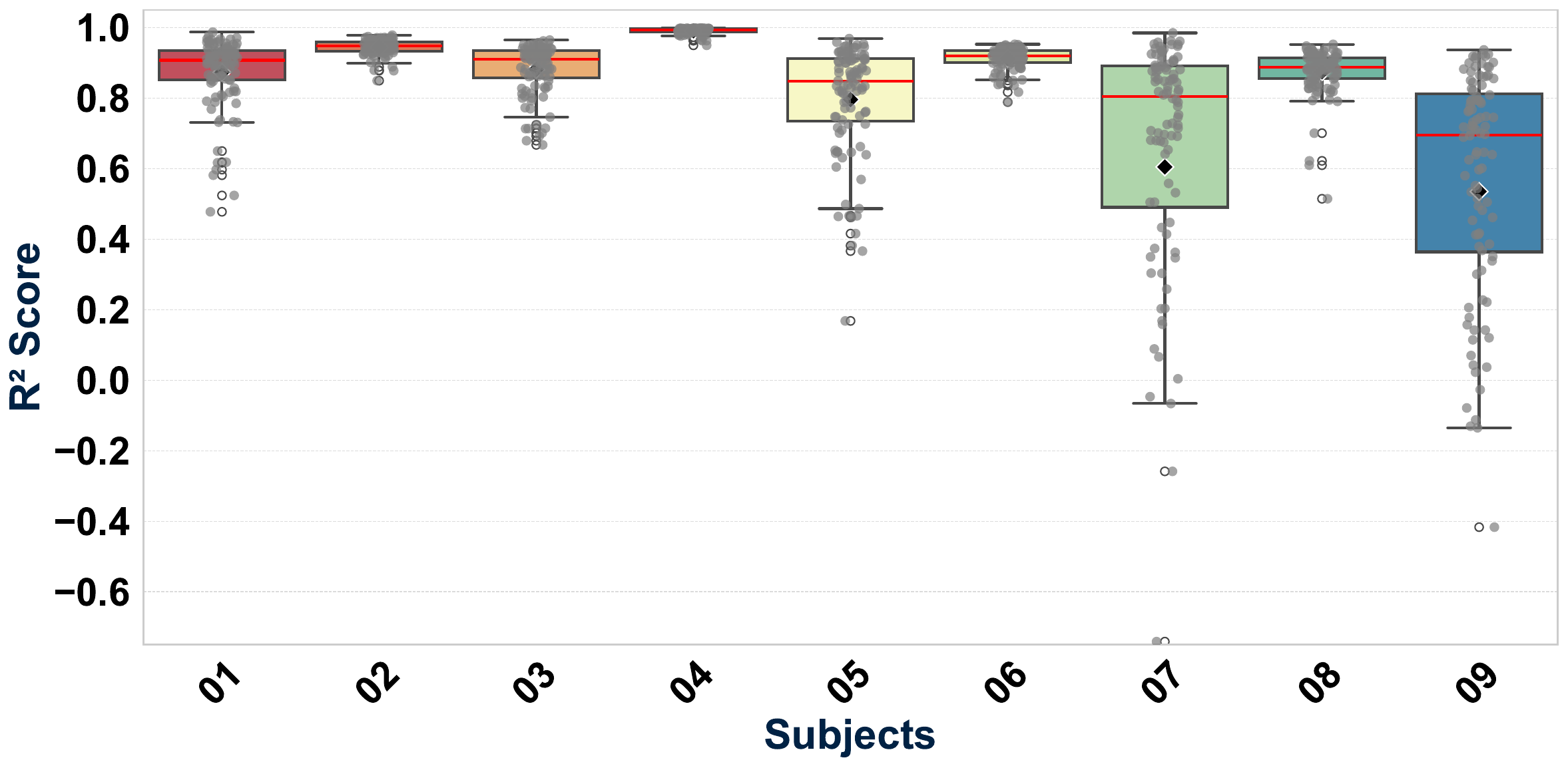}
            \caption{$R^2$ scores for the FastText embeddings.} 
            \label{fig:fasttext}  
    \end{figure}

    \begin{figure}[t]  
            \centering 
            \footnotesize
            \includegraphics[width=\linewidth]{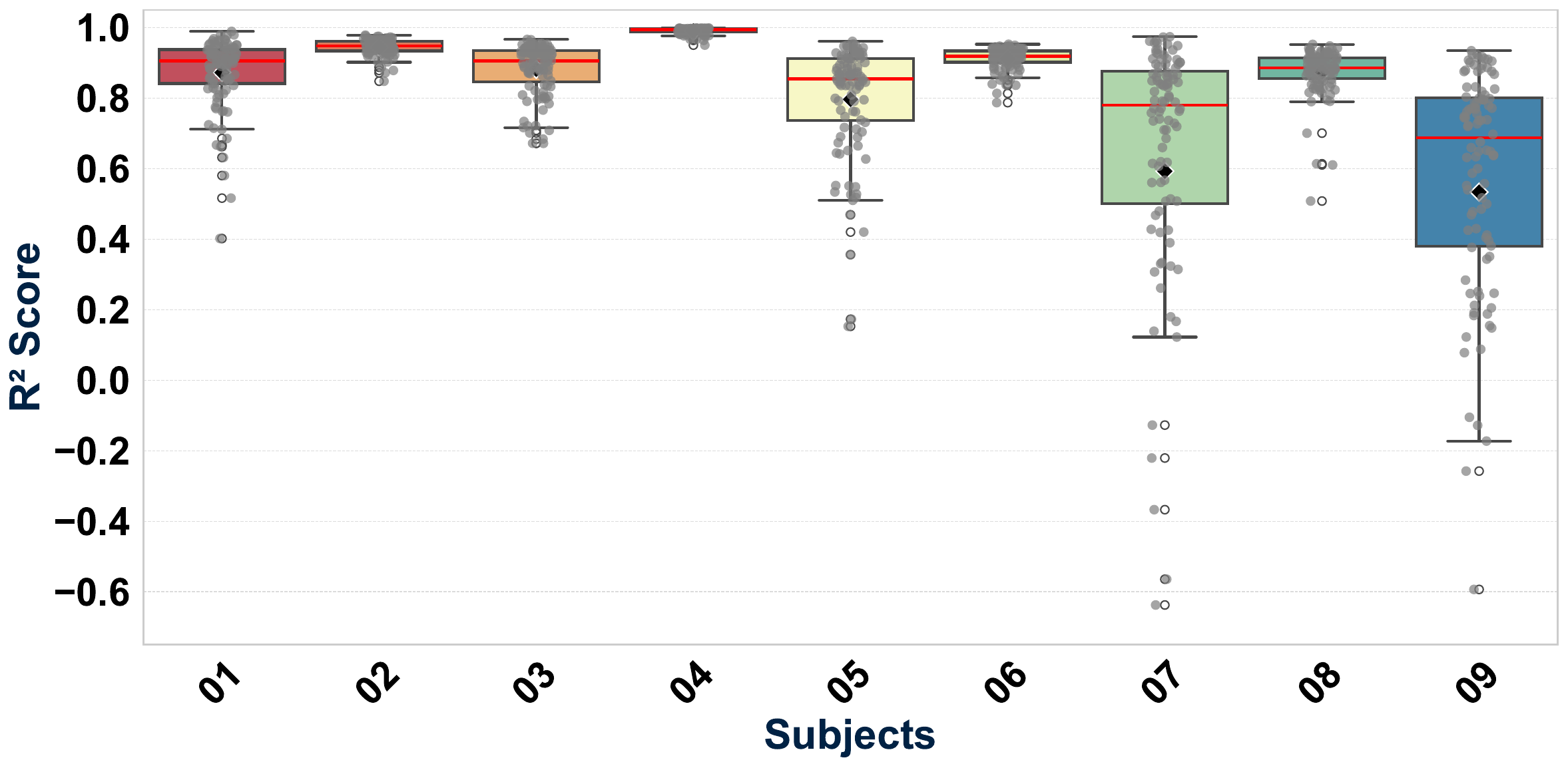}
            \caption{$R^2$ scores for the GPT-2.0 embeddings.} 
            \label{fig:chatgpt}  
    \end{figure}

        Figs. \ref{fig:fasttext} and \ref{fig:chatgpt} present the $R^2$ scores for the FastText and GPT-2.0 models, respectively. Both models exhibit similarities in their predictive performance, with certain subjects consistently achieving high $R^2$ values. Notably, subject S04 achieved near-perfect reconstruction, with a mean $R^2$ score of 0.99 for both FastText and GPT-2.0, suggesting that when neural data aligns well with the learned representations, both models can achieve exceptional accuracy. Conversely, subjects S05, S07, and S09 showed the lowest $R^2$ scores, ranging from 0.53 to 0.79, indicating lower predictive performance in these cases. Differences in the $R^2$ coefficients among participants can be attributed to variations in neural activity subject, largely influenced by the number and placement of implanted electrodes. In particular, S04 exhibits a greater density of electrodes in the Broca and Wernicke areas, that are brain regions strongly linked to speech and language processing. In constrast, S05, S07, and S09 have fewer electrodes implanted in those areas, which may hinder the ability to effectively monitor critical for language-related functions. This disparity in electrode distribution likely contributes to the variability observed across participants.
    
        While both models FastText and GPT-2 performed similarly overall, distinct differences emerged. In S07, FastText outperformed GPT-2.0, achieving an $R^2$ score of 0.60, suggesting that FastText embeddings may better capture certain linguistic structures or semantic relationships. Additionally, FastText demonstrated slightly better performance for subjects S06 to S09, whereas GPT-2.0 achieved marginally higher accuracy in subjects S01, S03, S04, and S05. These differences, though subtle, indicate potential variations in how each model encodes linguistic information relevant to neural activity.

    \subsection{Performance of Wav2Vec 2.0 embeddings}
        \begin{figure}[t]  
            \centering 
            \footnotesize
            \includegraphics[width=\linewidth]{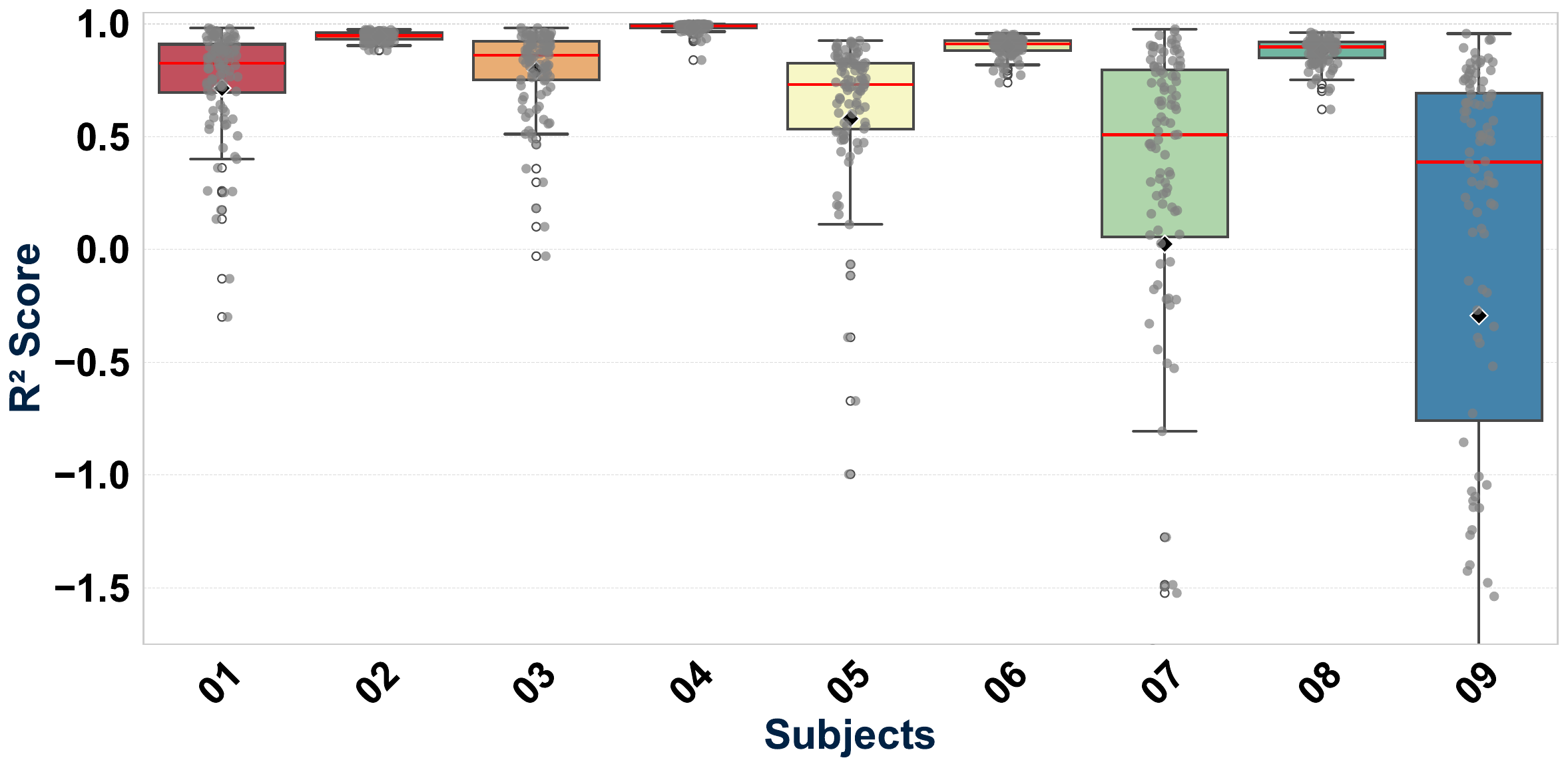}
            \caption{$R^2$ scores for the Wav2Vec 2.0 embeddings.} 
            \label{fig:Wav2vec}  
    \end{figure}
        Fig. \ref{fig:Wav2vec} shows the $R^2$ results across all subjects when using the Wav2Vec 2.0 embeddings. The highest $R^2$ values are observed in subjects S04 (0.98), S02 (0.94), and S06 (0.89), indicating strong predictive performance. Moderate reconstruction accuracy was achieved for subjects 03 (0.79), 08 (0.87), and 01 (0.71). However, for subjects S09 (-0.29) and S07 (0.02), the model failed to reconstruct neural activity accurately, suggesting that these embeddings may not be as effectively as the text-based ones to capture relevant features for neural reconstruction for our dataset.

    \subsection{Comparative analysis}
        \begin{figure}[t]  
            \centering 
            \footnotesize
            \includegraphics[width=\linewidth]{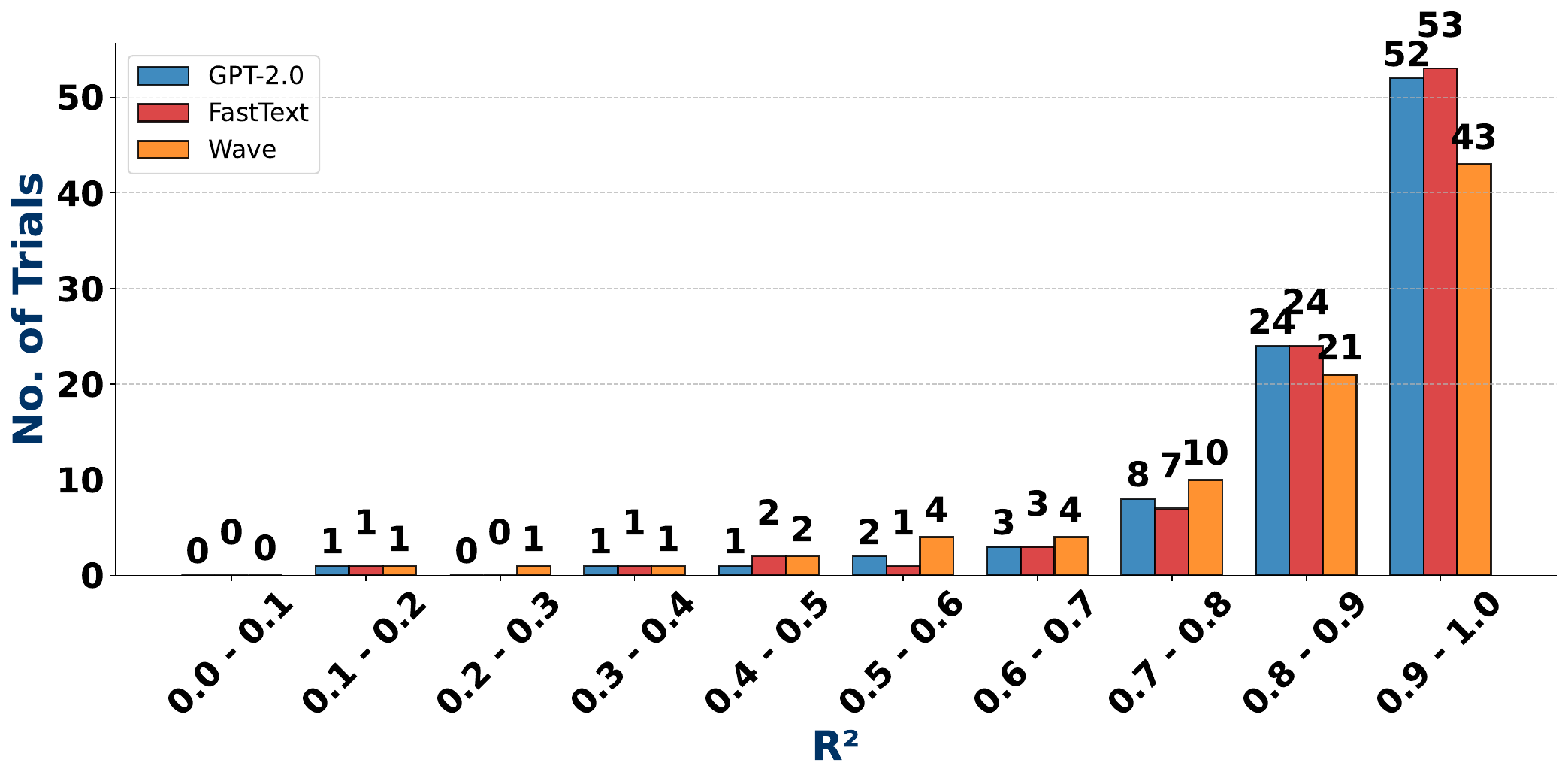}
            \caption{Average distribution of trials based on $R^2$ scores.} 
            \label{fig:comp_dist}  
    \end{figure}
        Fig. \ref{fig:comp_dist} shows the distribution of $R^2$ scores across all trials (words) for all subjects, comparing the three types of embeddings. The majority of trials yielded high $R^2$ scores, with FastText exhibiting the largest proportion of trials in the 0.9–1.0 range, followed by GPT-2.0 and Wav2Vec 2.0. This suggests that, on average, FastText embeddings offer slightly better neural reconstruction performance.

        \begin{table}[t]
            \centering
            \renewcommand{\arraystretch}{1.4}
            \setlength{\tabcolsep}{6pt} 
            \caption{Mean Pearson Correlation Coefficient (PCC) values across subjects for neural reconstruction using GPT-2.0, FastText, and Wav2Vec 2.0 embeddings. Standard deviation of PCC shown in parentheses. \label{tab:pcc_comparison}}
            \begin{tabular}{ c c c c}
                \toprule
                \textbf{Subject} & \textbf{GPT-2.0} & \textbf{FastText} & \textbf{Wav2Vec 2.0} \\
                \midrule
                S01 & 0.94 (0.01) & 0.94 (0.02) & 0.82 (0.33) \\
                S02 & 0.97 (0.02) & 0.97 (0.01) & 0.82 (0.02) \\
                S03 & 0.93 (0.02) & 0.93 (0.01) & 0.86 (0.16) \\
                S04 & 0.99 (0.01) & 0.99 (0.02) & 0.99 (0.02) \\
                S05 & 0.88 (0.01) & 0.88 (0.01) & 0.74 (0.49) \\
                S06 & 0.95 (0.01) & 0.95 (0.01) & 0.93 (0.04) \\
                S07 & 0.81 (0.02) & 0.81 (0.02) & 0.42 (0.59) \\
                S08 & 0.93 (0.02) & 0.93 (0.01) & 0.92 (0.06) \\
                S09 & 0.78 (0.05) & 0.79 (0.04) & 0.25 (0.69) \\
                \bottomrule
            \end{tabular}
        \end{table}

        Table \ref{tab:pcc_comparison} shows the neural reconstruction performance achieved in each case in terms of the PCC coefficient. Wav2Vec 2.0 exhibited the greatest variability, with mean PCC values ranging from 0.25 to 0.99, indicating that while it performed well for some subjects, its effectiveness was inconsistent. In contrast, both GPT-2.0 and FastText produced stable PCC values across subjects, suggesting a high degree of agreement in their predictive performance.

        In most subjects, the standard deviation of PCC remained low, indicating minimal variability across all trials. However, subject S09 showed a notable increase in standard deviation, particularly for GPT-2.0, suggesting possible inconsistencies in neural responses or differences in how the model represents certain linguistic features.

        Overall, these findings indicate that both FastText and GPT-2.0 reliably map linguistic stimuli to neural responses, with minimal differences in performance. The consistency of results across subjects underscores the robustness of these models in capturing the neural encoding of language. While slight variations in standard deviation exist, they do not significantly affect overall conclusions, reinforcing the effectiveness of both embedding approaches in studying neural representations of linguistic information.

\section{Conclusions}
    We explored the potential for speech neuroprostheses by leveraging embeddings from the pre-trained large-scale, self-supervised language and audio models to reconstruct the neural activity recorded during speech production. Our experiments provide evidence that the neural activity associated with speech production can be effectively modeled using machine learning and deep learning. The results show that both models achieved high PCCs. The overall trend suggests that neural embeddings from pre-trained models effectively capture speech-related brain activity, with consistent performance across participants. Future work will involve investigating how the integration of multimodal embeddings affects reconstruction accuracy. We will explore whether the transfer of information from lower-level to higher-level layers in deep learning models mirrors the neural information transfer observed in the brain, particularly in relation to the temporal dynamics of stimulus onset.

\section{Acknowledgments}
This work was supported by the grant PID2022-141378OB-C22 funded by MICIU/AEI/10.13039/501100011033 and ERDF/EU.
We are grateful to Dr. Christian Herff from Maastricht University for generously sharing the sEEG dataset with us.

\bibliographystyle{IEEEtran}
\bibliography{mybib}

\begin{thebibliography}{10}
\providecommand{\url}[1]{#1}
\csname url@samestyle\endcsname
\providecommand{\newblock}{\relax}
\providecommand{\bibinfo}[2]{#2}
\providecommand{\BIBentrySTDinterwordspacing}{\spaceskip=0pt\relax}
\providecommand{\BIBentryALTinterwordstretchfactor}{4}
\providecommand{\BIBentryALTinterwordspacing}{\spaceskip=\fontdimen2\font plus
\BIBentryALTinterwordstretchfactor\fontdimen3\font minus \fontdimen4\font\relax}
\providecommand{\BIBforeignlanguage}[2]{{%
\expandafter\ifx\csname l@#1\endcsname\relax
\typeout{** WARNING: IEEEtran.bst: No hyphenation pattern has been}%
\typeout{** loaded for the language `#1'. Using the pattern for}%
\typeout{** the default language instead.}%
\else
\language=\csname l@#1\endcsname
\fi
#2}}
\providecommand{\BIBdecl}{\relax}
\BIBdecl

\bibitem{Herff2015}
C.~Herff, D.~Heger, A.~De~Pesters, D.~Telaar, P.~Brunner, G.~Schalk, and T.~Schultz, ``Brain-to-text: decoding spoken phrases from phone representations in the brain,'' \emph{Frontiers in neuroscience}, vol.~9, p. 217, 2015.

\bibitem{Moses2021}
D.~A. Moses, M.~K. Leonard, J.~G. Makin, and E.~F. Chang, ``Neuroprosthesis for decoding speech in a paralyzed person with anarthria,'' \emph{New England Journal of Medicine}, vol. 385, no.~3, pp. 217--227, 2021.

\bibitem{Metzger2022}
S.~L. Metzger, J.~R. Liu, D.~A. Moses, M.~E. Dougherty, M.~P. Seaton, K.~T. Littlejohn, J.~Chartier, G.~K. Anumanchipalli, A.~Tu-Chan, K.~Ganguly \emph{et~al.}, ``Generalizable spelling using a speech neuroprosthesis in an individual with severe limb and vocal paralysis,'' \emph{Nature communications}, vol.~13, no.~1, p. 6510, 2022.

\bibitem{Willett2023}
F.~R. Willett, E.~M. Kunz, C.~Fan, D.~T. Avansino, G.~H. Wilson, E.~Y. Choi, F.~Kamdar, M.~F. Glasser, L.~R. Hochberg, S.~Druckmann \emph{et~al.}, ``A high-performance speech neuroprosthesis,'' \emph{Nature}, vol. 620, no. 7976, pp. 1031--1036, 2023.

\bibitem{Silva2024bilingual}
A.~B. Silva, J.~R. Liu, S.~L. Metzger, I.~Bhaya-Grossman, M.~E. Dougherty, M.~P. Seaton, K.~T. Littlejohn, A.~Tu-Chan, K.~Ganguly, D.~A. Moses \emph{et~al.}, ``A bilingual speech neuroprosthesis driven by cortical articulatory representations shared between languages,'' \emph{Nature Biomedical Engineering}, pp. 1--15, 2024.

\bibitem{Card2024}
N.~S. Card, M.~Wairagkar, C.~Iacobacci, X.~Hou, T.~Singer-Clark, F.~R. Willett, E.~M. Kunz, C.~Fan, M.~Vahdati~Nia, D.~R. Deo \emph{et~al.}, ``An accurate and rapidly calibrating speech neuroprosthesis,'' \emph{New England Journal of Medicine}, vol. 391, no.~7, pp. 609--618, 2024.

\bibitem{Anumanchipalli2019}
G.~K. Anumanchipalli, J.~Chartier, and E.~F. Chang, ``Speech synthesis from neural decoding of spoken sentences,'' \emph{Nature}, vol. 568, no. 7753, pp. 493--498, 2019.

\bibitem{Angrick2019}
M.~Angrick, C.~Herff, E.~Mugler, M.~C. Tate, M.~W. Slutzky, D.~J. Krusienski, and T.~Schultz, ``Speech synthesis from ecog using densely connected 3d convolutional neural networks,'' \emph{Journal of neural engineering}, vol.~16, no.~3, p. 036019, 2019.

\bibitem{Herff2019}
C.~Herff, L.~Diener, M.~Angrick, E.~Mugler, M.~C. Tate, M.~A. Goldrick, D.~J. Krusienski, M.~W. Slutzky, and T.~Schultz, ``Generating natural, intelligible speech from brain activity in motor, premotor, and inferior frontal cortices,'' \emph{Frontiers in neuroscience}, vol.~13, p. 1267, 2019.

\bibitem{Angrick2021}
M.~Angrick, M.~C. Ottenhoff, L.~Diener, D.~Ivucic, G.~Ivucic, S.~Goulis, J.~Saal, A.~J. Colon, L.~Wagner, D.~J. Krusienski \emph{et~al.}, ``Real-time synthesis of imagined speech processes from minimally invasive recordings of neural activity,'' \emph{Communications biology}, vol.~4, no.~1, p. 1055, 2021.

\bibitem{Angrick2024}
M.~Angrick, S.~Luo, Q.~Rabbani, D.~N. Candrea, S.~Shah, G.~W. Milsap, W.~S. Anderson, C.~R. Gordon, K.~R. Rosenblatt, L.~Clawson \emph{et~al.}, ``Online speech synthesis using a chronically implanted brain--computer interface in an individual with {ALS},'' \emph{Scientific reports}, vol.~14, no.~1, p. 9617, 2024.

\bibitem{Khanday2025}
O.~M. Khanday, J.~L. P{\'e}rez-C{\'o}rdoba, M.~Y. Mir, A.~A. Najar, and J.~A. Gonzalez-Lopez, ``Neuroincept decoder for high-fidelity speech reconstruction from neural activity,'' \emph{arXiv preprint arXiv:2501.03757}, 2025.

\bibitem{Metzger2023}
S.~L. Metzger, K.~T. Littlejohn, A.~B. Silva, D.~A. Moses, M.~P. Seaton, R.~Wang, M.~E. Dougherty, J.~R. Liu, P.~Wu, M.~A. Berger \emph{et~al.}, ``A high-performance neuroprosthesis for speech decoding and avatar control,'' \emph{Nature}, vol. 620, no. 7976, pp. 1037--1046, 2023.

\bibitem{Gonzalez2020}
J.~A. Gonzalez-Lopez, A.~Gomez-Alanis, J.~M. Martín~Doñas, J.~L. Pérez-Córdoba, and A.~M. Gomez, ``Silent speech interfaces for speech restoration: A review,'' \emph{IEEE Access}, vol.~8, pp. 177\,995--178\,021, 2020.

\bibitem{Luo2023}
S.~Luo, M.~Angrick, C.~Coogan, D.~N. Candrea, K.~Wyse-Sookoo, S.~Shah, Q.~Rabbani, G.~W. Milsap, A.~R. Weiss, W.~S. Anderson \emph{et~al.}, ``Stable decoding from a speech bci enables control for an individual with {ALS} without recalibration for 3 months,'' \emph{Advanced Science}, vol.~10, no.~35, p. 2304853, 2023.

\bibitem{Vansteensel2024}
M.~J. Vansteensel, S.~Leinders, M.~P. Branco, N.~E. Crone, T.~Denison, Z.~V. Freudenburg, S.~H. Geukes, P.~H. Gosselaar, M.~Raemaekers, A.~Schippers \emph{et~al.}, ``Longevity of a brain--computer interface for amyotrophic lateral sclerosis,'' \emph{New England Journal of Medicine}, vol. 391, no.~7, pp. 619--626, 2024.

\bibitem{Puffay2023}
C.~Puffay, B.~Accou, L.~Bollens, M.~J. Monesi, J.~Vanthornhout, H.~Van~hamme, and T.~Francart, ``Relating {EEG} to continuous speech using deep neural networks: a review,'' \emph{Journal of Neural Engineering}, vol.~20, no.~4, p. 041003, 2023.

\bibitem{Silva2024}
A.~B. Silva, K.~T. Littlejohn, J.~R. Liu, D.~A. Moses, and E.~F. Chang, ``The speech neuroprosthesis,'' \emph{Nature Reviews Neuroscience}, vol.~25, no.~7, pp. 473--492, 2024.

\bibitem{Chang2024}
E.~F. Chang, ``Brain–computer interfaces for restoring communication,'' \emph{New England Journal of Medicine}, vol. 391, no.~7, pp. 654--657, 2024.

\bibitem{Chen2024}
X.~Chen, R.~Wang, A.~Khalilian-Gourtani, L.~Yu, P.~Dugan, D.~Friedman, W.~Doyle, O.~Devinsky, Y.~Wang, and A.~Flinker, ``A neural speech decoding framework leveraging deep learning and speech synthesis,'' \emph{Nature Machine Intelligence}, pp. 1--14, 2024.

\bibitem{Herff2020}
C.~Herff, D.~J. Krusienski, and P.~Kubben, ``The potential of stereotactic-eeg for brain-computer interfaces: current progress and future directions,'' \emph{Frontiers in neuroscience}, vol.~14, p. 123, 2020.

\bibitem{Kohler2022}
J.~Kohler, M.~C. Ottenhoff, S.~Goulis, M.~Angrick, A.~J. Colon, L.~Wagner, S.~Tousseyn, P.~L. Kubben, and C.~Herff, ``Synthesizing speech from intracranial depth electrodes using an encoder-decoder framework,'' \emph{Neurons, Behavior, Data analysis, and Theory}, vol.~6, no.~1, dec 9 2022.

\bibitem{Angrick2022}
M.~Angrick, M.~Ottenhoff, L.~Diener, D.~Ivucic, G.~Ivucic, S.~Goulis, A.~J. Colon, L.~Wagner, D.~J. Krusienski, P.~L. Kubben \emph{et~al.}, ``Towards closed-loop speech synthesis from stereotactic {EEG}: {A} unit selection approach,'' in \emph{Proc. ICASSP}, 2022, pp. 1296--1300.

\bibitem{Baevski2020wav2vec}
A.~Baevski, Y.~Zhou, A.~Mohamed, and M.~Auli, ``{Wav2vec 2.0}: {A} framework for self-supervised learning of speech representations,'' \emph{Advances in neural information processing systems}, vol.~33, pp. 12\,449--12\,460, 2020.

\bibitem{Radford2019}
A.~Radford, J.~Wu, R.~Child, D.~Luan, D.~Amodei, I.~Sutskever \emph{et~al.}, ``Language models are unsupervised multitask learners,'' \emph{OpenAI blog}, vol.~1, no.~8, p.~9, 2019.

\bibitem{Millet2022}
J.~Millet, C.~Caucheteux, Y.~Boubenec, A.~Gramfort, E.~Dunbar, C.~Pallier, J.-R. King \emph{et~al.}, ``Toward a realistic model of speech processing in the brain with self-supervised learning,'' \emph{Advances in Neural Information Processing Systems}, vol.~35, pp. 33\,428--33\,443, 2022.

\bibitem{Tuckute2023}
G.~Tuckute, J.~Feather, D.~Boebinger, and J.~H. McDermott, ``Many but not all deep neural network audio models capture brain responses and exhibit correspondence between model stages and brain regions,'' \emph{Plos Biology}, vol.~21, no.~12, p. e3002366, 2023.

\bibitem{Chen2024self}
P.~Chen, L.~He, L.~Fu, L.~Fan, E.~F. Chang, and Y.~Li, ``Do self-supervised speech and language models extract similar representations as human brain?'' in \emph{Proc. ICASSP}, 2024, pp. 2225--2229.

\bibitem{tay2023elastic}
J.~K. Tay, B.~Narasimhan, and T.~Hastie, ``Elastic net regularization paths for all generalized linear models,'' \emph{Journal of statistical software}, vol. 106, 2023.

\bibitem{fasttext}
\BIBentryALTinterwordspacing
A.~Joulin, E.~Grave, P.~Bojanowski, M.~Douze, H.~Jégou, and T.~Mikolov, ``Fasttext.zip: Compressing text classification models,'' 2016. [Online]. Available: \url{https://arxiv.org/abs/1612.03651}
\BIBentrySTDinterwordspacing

\bibitem{Verwoert2022}
M.~Verwoert, M.~C. Ottenhoff, S.~Goulis, A.~J. Colon, L.~Wagner, S.~Tousseyn, J.~P. Van~Dijk, P.~L. Kubben, and C.~Herff, ``Dataset of speech production in intracranial electroencephalography,'' \emph{Scientific data}, vol.~9, no.~1, p. 434, 2022.

\bibitem{ifaCorpus}
R.~van Son, D.~Binnenpoorte, and L.~Pols, ``The {IFA} corpus: A phonemically segmented dutch ‘open source’ speech database,'' in \emph{Proceedings of Eurospeech}.\hskip 1em plus 0.5em minus 0.4em\relax Data Archiving and Networked Services (DANS), Sep 2001.

\bibitem{babu2021xls}
A.~Babu, C.~Wang, A.~Tjandra, K.~Lakhotia, Q.~Xu, N.~Goyal, K.~Singh, P.~Von~Platen, Y.~Saraf, J.~Pino \emph{et~al.}, ``{XLS-R}: Self-supervised cross-lingual speech representation learning at scale,'' \emph{arXiv preprint arXiv:2111.09296}, 2021.

\end{thebibliography}

\end{document}